\documentclass[aps,prd,twocolumn,groupedaddress,showpacs]{revtex4-1}
\usepackage{amsfonts,amsmath,amssymb}
\begin{document}

\title{Magnetic susceptibility of the quark condensate via holography}

\author{A. Gorsky}
\email{gorsky@itep.ru}
\affiliation{Institute for Theoretical and Experimental
Physics \\ B. Cheremushkinskaya ul. 25, 117259
Moscow, Russia}

\author{A. Krikun   }
\email{krikun.a@gmail.com}
\affiliation{Institute for Theoretical and Experimental
Physics \\ B. Cheremushkinskaya ul. 25, 117259
Moscow, Russia}

\preprint{ITEP-TH-04/09}

\date{\today}

\begin{abstract}
We discuss the holographic derivation of the magnetic susceptibility
of the quark condensate. It is found that the susceptibility emerges upon
the account of the Chern-Simons term in the  holographic action. We
demonstrate that  Vainshtein's relation is not exact in the hard wall dual model
but is fulfilled with  high accuracy.
Some comments concerning the spectral density of the Dirac operator are presented.

\end{abstract}

\pacs{ 11.25.Tq, 11.40.Ha, 11.15.Ex, 12.38.Aw }

\maketitle

\multlinegap=0 in

\newcommand{\p}{\partial}
\newcommand{\ph}{\varphi}
\newcommand{\be}{\begin{equation}}
\newcommand{\ee}{\end{equation}}
\newcommand{\ba}{\begin{align}}
\newcommand{\ea}{\end{align}}
\newcommand{\bg}{\begin{gather}}
\newcommand{\eg}{\end{gather}}
\newcommand{\dz}{\partial_{z}}
\newcommand{\dx}{\partial_{x}}

\section{Introduction}
The analysis of the QCD properties by holographic methods is one
of the most promising approaches to the description of the strong
coupling region. The unique holographic model for QCD has not been found yet
hence there is no hope to get the generic quantitative predictions at present.
However there are some QCD results which seem to be independent on the
details of the dual geometry hence one could consider these universal
objects or relations to test the holographic picture. On the other
hand it is instructive to analyze  if some relation is universal indeed
testing it in the different holographic geometries.

The simplest relation to be tested is the Gell-Mann-Rennes-Oaks one
which was shown to be true in all holographic models of QCD like
hard wall models \cite{Erlich, Westenberger}, soft wall model \cite{soft} or Sakai-Sugimoto
model \cite{Sugimoto}. The main focus in our paper is the magnetic susceptibility
of the quark condensate describing the response of the QCD vacuum
on the external magnetic field. It was introduced in \cite{is}  in the context of the
sum rules  and investigation of its numerical value was performed in \cite{bal,bel,dorokhov1,dorokhov2}.
More recently using QPE arguments Vainshtein  \cite{Vain}  obtained the expression for the
susceptibility in terms of the known QCD quantities .
However the status of this relation is questionable since both sides of the
correspondence have different anomalous dimensions and it is not clear if
the higher  states could influence the answer. (see \cite{Braun, Rohr})

In this paper we shall analyze the magnetic susceptibility
in the holographic setting and shall focus mostly at the simplest
hard wall model \cite{Erlich,Pomarol}(introduced in \cite{Polchinski,Boschi1,Boschi2}). We shall consider the calculation of the
three-point function similar the two-point calculations in \cite{Pomarol,Krik}
and formfactor calculations in \cite{Rad1,Rad2}. We shall consider the special
kinematics of the three-point function related to the
susceptibility. It turns out that the only nontrivial contribution
to the correlator comes from the Chern-Simons term in the dual action
and substituting  the solutions to the classical equations of motion
we get the result for the susceptibility which is close
to the Vainshtein's relation.

The paper is organized as follows. First in Section II we calculate the three-point function and the magnetic susceptibility of quark condensate in the AdS/QCD hard wall model. In Section III we survey some other approaches to the calculation of this value, namely via Chiral Perturbation theory and via the relation with the Dirac operator spectrum density. The conclusion is given in Section IV. To make the paper self consistent, we state in the Appendix some results of \cite{Erlich,Krik}, which we will use.

\section{ Hard wall AdS/QCD model.}
\subsection {Chern-Simons action and 3-point function}

Our aim is to calculate the correlation function of two vector and one axial  currents.
To make the holographic calculation we take the simple "hard wall" AdS/QCD model \cite{Erlich,Pomarol}. We will work in notation of \cite{Krik} and use some results, calculated in \cite{Erlich,Krik} (see also Appendix).
The holographic action involves kinetic and
Chern-Simons term
:
$$
S= S_{YM}(A_L.A_R)  +  S_{CS}(A_L)-S_{CS}(A_R)
$$
where
$$
S_{CS}(A)=\frac{N_C}{24 \pi^2} \int Tr \left(AF^2-\frac{1}{2} A^3 F + \frac{1}{10} A^5 \right)
$$
It is clear that only terms containing 3 gauge fields ($A,V,V$)are relevant for the calculation
of correlator:
\begin{multline*}
Tr(A_L F_L F_L - A_R F_R F_R) \rightarrow \\
 \xrightarrow[AVV]{}  2 Tr (V F_V F_A + V F_A F_V + A F_V F_V)
\end{multline*}

The classical solutions for the fields have the form:
\begin{gather}
V_{\mu}^{a}(z,Q) = \hat{V}_{\nu}^{a}(Q) \cdot V_{\mu \nu}(Q^2,z) \label{sources}\\
A_{\mu}^{a}(z,Q) = \hat{A}_{\nu}^{a}(Q) \cdot A_{\mu \nu}(Q^2,z) \notag\\
A_{\mu \nu}(Q^2,z) = P(Q)^{\perp}_{\mu \nu} a_{\perp}(Q^2,z) + P(Q)^{\parallel}_{\mu \nu} a_{\parallel}(Q^2,z) \notag \\
V_{\mu \nu}(Q^2,z) = P(Q)^{\perp}_{\mu \nu} v(Q^2,z) \notag
\end{gather}
where $\hat{V}_{\nu}^{a}(Q)$ and $\hat{A}_{\nu}^{a}(Q)$ provide the sources for
the operators. We assume the currents to have arbitrary charges with respect to $SU(2)$ group,
so the gauge group structure of the result is:

\begin{multline*}
\langle A V \tilde{V} \rangle= \frac{\delta^3}{\delta \hat{A} \delta \hat{V} \delta \hat{\tilde{V}}}  S_{CS} = \\
\begin{split}
=\frac{N_C}{12 \pi^2}  \int &\langle T_A T_V T_{\tilde{V}} \rangle
\cdot (V \tilde{F_V} F_A + \tilde{V} F_A F_V + A F_V \tilde{F_V}) \\
+ &\langle T_A T_V T_{\tilde{V}} \rangle \cdot (\tilde{V} F_V F_A + V F_A \tilde{F_V} + A \tilde{F_V} F_V)
\end{split}
\end{multline*}

We will work in the gauge $A_z = V_z = 0$. To figure out the integral over $z$, we rewrite it as:
\begin{multline*}
\int  V \tilde{F_V} F_A + \tilde{V} F_A F_V + A F_V \tilde{F_V}= \\
\begin{split}
  =4 \epsilon^{\mu \nu \rho \sigma} \int &  \p_z A_{\mu} [\p_{\sigma} V_{\nu} \tilde{V}_{\rho} -  V_{\nu} \p_{\sigma} \tilde{V}_{\rho}] \\
 + &A_{\mu}[ \p_z V_{\nu} \p_{\sigma} \tilde{V}_{\rho} - \p_{\sigma} V_{\nu} \p_z \tilde{V}_{\rho}] \\
 + &\p_{\sigma} A_{\mu} [ V_{\nu} \p_z \tilde{V}_{\rho} - \p_z V_{\nu} \tilde{V}_{\rho}]
\end{split}
\end{multline*}

Substituting  Fourier components of fields (\ref{sources}) and introducing the tensors
\begin{align*}
\epsilon^{\perp}_{\mu \nu \rho \sigma} &= \epsilon^{\alpha \beta \gamma \sigma} P^{\perp}_{\alpha \mu}(k_1)P^{\perp}_{\beta \nu}(k_2)P^{\perp}_{\gamma \rho}(k_3) \\
\epsilon^{\parallel}_{\mu \nu \rho \sigma} &= \epsilon^{\alpha \beta \gamma \sigma} P^{\parallel}_{\alpha \mu}(k_1)P^{\perp}_{\beta \nu}(k_2)P^{\perp}_{\gamma \rho}(k_3)
\end{align*}
we get (denote $\tilde{v} =v(k_3)$, $\dot{v}= \p_z v$) :
\begin{multline*}
\langle A_{\perp \mu}(k_1) V_{\nu}(k_2) \tilde{V}_{\rho}(k_3) \rangle= \\
\begin{split}
= \frac{N_C}{12 \pi^2} \langle T_A T_V T_{\tilde{V}} \rangle  \ & 4 \delta^4(k_1 + k_2 + k_3) \epsilon^{\perp}_{\mu \nu \rho \sigma}  \\
\times \int dz  \ (i{k_2}_{\sigma})& [ \dot{a}_{\perp}v \tilde{v} + a_{\perp} \dot{v} \tilde{v} - 2 a_{\perp} v \dot{\tilde{v}}]  \\
-(i{k_3}_{\sigma})& [ \dot{a}_{\perp}v \tilde{v} + a_{\perp} v \dot{\tilde{v}} -2 a_{\perp} \dot{v} \tilde{v}]
\end{split}
\end{multline*}
We can add a surface term $(-ik_2+ik_3)\p_z(a_{\perp}v \tilde{v})$ in the action, in order to make the 3-point function vanish, if one of vector momenta tends to zero. This will lead us to the expression:

\begin{widetext}

\begin{align}
\langle A_{\perp \mu}(k_1) V_{\nu}(k_2) \tilde{V}_{\rho}(k_3) \rangle=- \frac{N_C}{\pi^2}
\langle T_A T_V T_{\tilde{V}} \rangle  \  \delta^4(k_1 + k_2 + k_3) \epsilon^{\perp}_{\mu \nu \rho \sigma} \int dz  \  (i{k_2}_{\sigma}) a v \dot{\tilde{v}} - (i{k_3}_{\sigma})a \dot{v} \tilde{v}
\end{align}
Similarly
\begin{align}
\label{3PF}
\langle A_{\parallel \mu}(k_1) V_{\nu}(k_2) \tilde{V}_{\rho}(k_3) \rangle=- \frac{N_C}{\pi^2}
\langle T_A T_V T_{\tilde{V}} \rangle \ \delta^4(k_1 + k_2 + k_3) \epsilon^{\parallel}_{\mu \nu \rho \sigma} \int dz  \  (i{k_2}_{\sigma}) a v \dot{\tilde{v}} - (i{k_3}_{\sigma})a \dot{v} \tilde{v}
\end{align}

\end{widetext}

\subsection {Solution for $A_{\perp}$}
Let us consider the equation of motion for $A_{\perp}$  at small $Q^2$ similarly to calculation in \cite{Rad1}.  The equation is \cite{Krik}:
$$
\left [ \partial_z \left ( \frac{1}{z} \partial_z A_{\mu}^{a} \right ) + \frac{q^2}{z} A_{\mu}^{a} - \frac{R^{2} g_{5}^{2} \Lambda^2 v^{2}}{z^{3}} A_{\mu}^{a} \right ]_\perp = 0
$$
We will work with the bulk-to-boundary propagator $a_{\perp}(z)$, which is defined in (\ref{sources}) and denote $R^{2} g_{5}^{2} \Lambda^2 = k^2= 3$ (\ref{k^2}). Using variable $y=\frac{k \sigma}{3} z^3=\alpha^3 z^3$, we get the equation
\begin{multline*}
\p_{y}^2 a + \frac{1}{3y} \p_y a - a = \frac{Q^2}{9 \alpha^2} y^{-4/3} a \\ + \frac{2 k^2 m \sigma}{9 \alpha^4} y^{-2/3} a + \frac{m^2}{9 \alpha^2} y^{-4/3} a
\end{multline*}
 which is an inhomogeneous modified Bessel equation. We introduce here the dimension parameter $\alpha$ which equals $\left(\frac{k \sigma}{3}\right)^{1/3} = 395 Mev$ (see (\ref{sigma}),(\ref{k^2})).    One can argue that the last term is negligible and
 solution to homogeneous part is
\begin{equation}
\label{a_perp}
a^{(0)}(y)=F y^{1/3} [ AI_{1/3}(y) + BK_{1/3}(y)],
\end{equation}
where constants are fixed by the conditions on the IR boundary $y_m=\alpha^3 z_{m}^3=1.82$ (see (\ref{z_m})):
\begin{gather*}
\p_z a(z)|_{z=z_m} = 3 \alpha y^{2/3} \p_y a(y)|_{y_m} = 0 \\
A=K_{2/3}(y_m); \qquad B=  I_{-2/3}(y_m)
\end{gather*}
and UV boundary:
\begin{gather*}
a(z)|_{z=\epsilon} = F y^{1/3}  \ B y^{-1/3} \frac{\Gamma(1/3)}{2^{2/3}}=1 \\
F = \frac{2^{2/3}}{B\Gamma(1/3)} .
\end{gather*}

Given this solution (which corresponds to Q=0), we can compute $f_{\pi}$, using the recipe, described in \cite{Erlich} (see (\ref{g_5})).
\begin{align}
\label{f_pi}
f_{\pi}^2 &= -\frac{R}{g_5^2} \frac{\p_z a(z)}{z}|_{z=0,Q=0} = \frac{R}{g_5^2} 1.815  \ \alpha^2 \sim (85 Mev)^2
\end{align}

\subsection {Solution for $A_\parallel$}

To obtain the longitudinal part of the 3-point function  we need to find bulk-to-boundary propagator in the pseudoscalar sector. It is the solution to equations \cite{Krik}:
\begin{equation}
\label{9}
 \partial_z \left ( \frac{1}{z} \partial_z \varphi^{a} \right ) + \frac{R^{2} g_{5}^{2} v^{2}}{z^{3}} (\pi^{a}- \varphi^{a})= 0
\end{equation}
\begin{equation}
\label{9.5}
Q^2 \partial_z \varphi^{a} + \frac{R^{2} g_{5}^{2} v^{2}}{z^{2}} \partial_z \pi^{a} = 0,
\end{equation}
where $\ph$ is related to the longitudinal part of $A_\mu$ as $A_{\parallel \mu} = \p_\mu \ph$. We introduce the function $\psi(z) = \ph(z) - \pi(z)$, and eliminate $\pi(z)$ from the system to get an equation on $\psi$ with the dimensionless variable $t= \alpha z$
\begin{equation}
\notag
t \partial_t \left ( \frac{1}{t} \partial_t \psi \right ) - \frac{k^{2} v^{2}}{t^{2}} \psi - t \p_t \left(\frac{1}{t} \frac{q}{\frac{k^{2} v^{2}}{t^{2}} + q}  \partial_t \psi  \right)= 0,
\end{equation}
where $q=Q^2/\alpha^2$. Now we can substitute $v(t) = \frac{\sigma}{\alpha^3} t^3 + \frac{m}{\alpha} t$ and write down terms up to the first order in $m/\alpha$ and $q$, assuming $Q^2$ to be small enough.
\begin{multline}
\label{eq_psi(y)}
 \p_y^2 \psi + \frac{1}{3} \frac{\p_y \psi}{y} - \psi =\frac{2}{9} \frac{k^2 m \sigma}{\alpha^4} y^{-2/3} \psi   \\  +\frac{q}{9} y^{-4/3} \left[1 - \frac{4}{3} \frac{\p_y}{y}\right] \psi + O\left(q^2,q \frac{m \sigma}{\alpha^4}\right),
\end{multline}
where $y=t^3$.

The homogeneous solution, subject to the boundary conditions $\pi(\epsilon) = 0 , \ph(\epsilon) =1, \p_z \pi(z_m)= \p_z \ph(z_m)=0$ is the same as for $a_{\perp}$ (\ref{a_perp}) as expected at $Q^2=0$. The Green function of this equation is:
\begin{align*}
G(u,v) = \frac{u^{1/3} v^{1/3}}{AD-BC} &[ AI_{1/3}(u) + BK_{1/3}(u)]
\\ \times &[ CI_{1/3}(v) + DK_{1/3}(v)]
\end{align*}
with $C$ and $D$ defined by the condition:
\begin{gather*}
G(y,y')|_{y=\epsilon} = 0 \\
C= -K_{1/3}(\epsilon); \qquad D= I_{1/3}(\epsilon).
\end{gather*}
It satisfies the equation
$$
\left[\p_{y}^2 + \frac{1}{3y} \p_y - 1\right]G(y,y') = \delta(y-y') \frac{1}{y^{1/3}}.
$$

We can compute the correction due to the quark mass in (\ref{eq_psi(y)}). It is obtained by the integral:

\begin{widetext}

\begin{align*}
\psi^{(m)}(y)&= \frac{2 k^2 m \sigma}{9 \alpha^4} \int_{\epsilon}^{y_m} \ y'^{1/3} G(y,y') \ y'^{-2/3} \psi^{(0)}(y')=\\
&= \frac{2 k^2 m \sigma}{9 \alpha^4} \frac{Fy^{1/3}[ AI_{1/3}(y) + BK_{1/3}(y)]}{AD-BC} \int_{\epsilon}^{y} \ y'^{1/3} [ CI_{1/3}(y') + DK_{1/3}(y')][ AI_{1/3}(y') + BK_{1/3}(y')]\\
&+ \frac{2 k^2 m \sigma}{9 \alpha^4} \frac{Fy^{1/3}[ CI_{1/3}(y) + DK_{1/3}(y)]}{AD-BC} \int_{y}^{y_m} \ y'^{1/3} [ AI_{1/3}(y') + BK_{1/3}(y')][ AI_{1/3}(y') + BK_{1/3}(y')]
\end{align*}

\end{widetext}

Using the value of $y_m=1.82$, which corresponds to the IR boundary in our model, we get for the correction:
$$
\psi^{(m)}(y) = - 1.0004 \cdot \frac{2k^2 m \sigma}{9\alpha^2} z^2
$$

Consequently, the solution for $\ph$ with correction due to the quark mass is:
\begin{multline}
\label{solution_ph}
\ph(z) = \psi(z) + O(q) =  \\ = F \alpha z  [ AI_{1/3}(y) + BK_{1/3}(y)] - \frac{2}{9}\frac{k^2 m \sigma}{\alpha^2} z^2
\end{multline}
we neglect here the correction due to the q term assuming that  $Q \ll 3\alpha = 1180 Mev$.

\subsection {Magnetic susceptibility of the quark condensate}
In this Subsection we calculate the magnetic susceptibility $\chi$ of the quark condensate defined as
\be
\langle \bar{q} \sigma_{\mu\nu}q \rangle_{F}= \chi \langle \bar{q}q \rangle F_{\nu\mu}
\ee
In order to find magnetic susceptibility , we study the 3-point function
$$\langle A_{\parallel \mu}(-Q) V_{\nu}(Q-k_3) \tilde{V}_{\rho}(k_3) \rangle$$     in the limit
$k_3\rightarrow 0$, according to \cite{Vain} where the following expression for the
susceptibility has been obtained
\be
\label{Chi}
\chi =- 2 \frac{N_c}{8 \pi^2} \frac{1}{f_{\pi}^2}.
\ee

Consider the classical solutions for the vector fields, calculated in \cite{Erlich,Krik}.
\begin{align*}
v(Q,z) &= Qz \left( K_{1}(Qz) + \frac{K_{0}(Qz_m)}{I_{0}(Qz_m)} I_{1}(Qz) \right)\xrightarrow[Q \rightarrow 0]{} 1\\
\dot{v}(k,z) &= k^2 z \left( K_{0}(kz) + \frac{K_{0}(kz_m)}{I_{0}(kz_m)} I_{0}(kz) \right)\sim k^2 ln(k)
\end{align*}
and substitute them into the correlator  (\ref{3PF}) :

\begin{multline*}
\langle A_{\parallel \mu}(-Q) V_{\nu}(Q-k_3) \tilde{V}_{\rho}(k_3) \rangle= \\
\begin{split}
= \frac{N_C}{\pi^2} \langle T_A T_V T_{\tilde{V}} \rangle  \ & \epsilon^{\parallel}_{\mu \nu \rho \sigma}(i{k_3}_{\sigma}) \int dz  \ a_{\parallel}(z) \dot{v}(Q,z) = \\
= \frac{N_C}{\pi^2}  \langle T_A T_V T_{\tilde{V}} \rangle \ & \epsilon^{\parallel}_{\mu \nu \rho \sigma}(i{k_3}_{\sigma}) \\
 \times \int dz &\left[\ph^{(0)}(z)+ \ph^{(m)}(z) \right] [Q^2 z K_{0}(Qz)]
\end{split}
\end{multline*}

In this integral due to the fast fall of the vector propagator $K_0(Qz)$ we can take the boundary value of $\phi^{(0)}$ in the first term. The second term can be  calculated  explicitly
\begin{equation*}
\int dz \ph^{(0)}(z) [Q^2 z K_{0}(Qz)] = \ph^{0}(0) \int dz Q^2 z K_{0}(Qz) =  \frac{1}{2}
\end{equation*}
\begin{multline*}
\int dz \ph^{(m)}(z) [Q^2 z K_{0}(Qz)] = \\ = \frac{2}{9} \frac{k^2 m \sigma}{\alpha^2} \int dz z^2 [Q^2 z K_{0}(Qz)] =   1.075  \ \frac{m \langle \bar{q} q \rangle}{Q^2 f_\pi^2}
\end{multline*}
where we've used the result (\ref{f_pi}) and the relation (\ref{sigma}). Finally, we get the expression for 3-point function with corrections:
\begin{widetext}
\begin{align*}
\langle A_{\parallel \mu}(-Q) V_{\nu}(Q-k_3) \tilde{V}_{\rho}(k_3) \rangle=  \langle T_A T_V T_{\tilde{V}} \rangle  \  \epsilon^{\parallel}_{\mu \nu \rho \sigma}(i{k_3}_{\sigma}) \left[ \frac{N_C}{2 \pi^2} - 1.075 \frac{N_c}{\pi^2} \frac{ m \langle \bar{q}q \rangle}{Q^2 f_{\pi}^2} + O(1/Q^4)\right]
\end{align*}
which can be matched the OPE of \cite{Vain}:
\begin{align*}
\langle A_{\parallel \mu}(-Q) V_{\nu}(Q) \tilde{V}_{\rho}(0) \rangle=  \langle T_A T_V T_{\tilde{V}} \rangle  \ \epsilon^{\parallel}_{\mu \nu \rho \sigma}(i{k_3}_{\sigma}) \left[  \frac{N_C}{2 \pi^2}   + \frac{4 m_f   \langle \bar{q}q \rangle  \chi}{Q^2} +O(1/Q^4)\right]
\end{align*}
\end{widetext}
This comparison allows us to determine the magnetic susceptibility of quark condensate $\chi$
\begin{equation}
\label{chi}
\chi =- 2.15 \frac{N_c}{8 \pi^2} \frac{1}{f_{\pi}^2}
\end{equation}
in close agreement with the result of Vainshtein (\ref{Chi}). This agreement is parametrical, but not numerical, because due to the small value of $f_\pi$(\ref{f_pi}) in our model, we get $\chi_{mod} = 11.5 \ \mathrm{Gev}^{-2}$ which is too large. Anyway, tuning the parameters of the model (mainly $\langle \bar{q}q \rangle$) can help fix $f_\pi$ to its real value and get the reasonable numerical agreement with Vainshtein's $\chi_{vain} = 8.9 \ \mathrm{Gev}^{-2}$. We've checked, that such variation of parameters do not affect coefficient in (\ref{chi}) significantly. Namely, its value changes less than 5\%, then the parameter $y_m$ which is proportional to ${\langle \bar{q}q \rangle} / {m_\rho^3}$ is varied in the wide range \hbox{from 1 to 8}.

\section{ Other approaches}
\subsection{3-point function in ChPT}

We can compute the same 3-point function in the Chiral Perturbation theory \cite{ChPT} and compare the result with AdS/QCD.
Note that the chiral Lagrangian is derived in the Sakai-Sugimoto \cite{Sugimoto} model hence the comment below can be considered
as the justification of the Vainshtein \cite{Vain} relation.
To obtain the $\langle AVV \rangle$ correlator, we  consider the Wess-Zumino-Witten term in ChPT action and turning on axial and vector external currents,

\begin{align*}
Z_{\chi PT} = \int d^4 x L_{2} + Z_{WZW}
\end{align*}
\begin{align*}
L_{2} =& \frac{F^2}{4} \langle D_{\mu}U D^{\mu} U^{\dag} \rangle,
\end{align*}
with $D_{\mu} = \p_{\mu} U - i r_{\mu} U + i U l_{\mu}$

\begin{multline*}
Z[U,l,r]_{WZW} =\\
\begin{split}
=& - \frac{i N_c}{240 \pi^2} \int_{M^5} d^5 x \epsilon^{ijklm} \langle \Sigma^{L}_{i}\Sigma^{L}_{j}\Sigma^{L}_{k}\Sigma^{L}_{l}\Sigma^{L}_{m} \rangle\\
& - \frac{i N_c}{48 \pi^2} \int d^4 x \epsilon_{\mu \nu \alpha \beta} \left( W(U,l,r)^{\mu \nu \alpha \beta} - W(1,l,r)^{\mu \nu \alpha \beta} \right)
\end{split}
\end{multline*}
\begin{multline*}
W(U,l,r)_{\mu \nu \alpha \beta} = \\
\begin{split}
&= \langle U l_\mu l_\nu l_\alpha U^{\dag} r_\beta + \frac{1}{4} U l_\mu U^\dag r_\nu U l_\alpha U^\dag r_\beta + i U \p_\mu l_\nu l_\alpha U^\dag r_\beta \\
& + i \p_\mu r_\nu U l_\alpha U^\dag r_\beta - i \Sigma^L_\mu l_\nu U^\dag r_\alpha U l_\beta + \Sigma^L_\mu U^\dag \p_\nu r_\alpha U l_\beta \\
&-\Sigma^L_\mu \Sigma^L_\nu U^\dag r_\alpha U l_\beta + \Sigma^L_\mu l_\nu \p_\alpha l_\beta + \Sigma^L_\mu \p_\nu l_\alpha l_\beta\\
&-i\Sigma^L_\mu l_\nu l_\alpha l_\beta + \frac{1}{2} \Sigma^L_\mu l_\nu \Sigma^L_\alpha l_\beta - i\Sigma^L_\mu \Sigma^L_\nu \Sigma^L_\alpha l_\beta \rangle\\
&-(L \leftrightarrow R).
\end{split}
\end{multline*}

Here
\begin{align*}
U=exp \left(\frac{i \sqrt{2}}{F} \pi^a t^a \right), \quad \Sigma_\mu^L = U^\dag \p_\mu U, \quad \Sigma_\mu^R = U \p_\mu U ^\dag
\end{align*}
and $(L \leftrightarrow R)$ stands for
$$
U \leftrightarrow U^{\dag} , \qquad l_\mu \leftrightarrow r_\mu , \qquad \Sigma_\mu^L \leftrightarrow \Sigma_\mu^R
$$

The leading contribution is given by the tree diagram, including $(a\pi)$ vertex from $L_{2}$ and $(\pi v v)$ vertex from $L_{WZW}$
\begin{multline*}
\begin{split}
 \Delta L_{2}[a \pi] =& \sqrt{2} F \langle \p_\mu \pi a^{\mu}\rangle \\
 \Delta L_{WZW}[\pi v v] =&  \frac{N_c}{48 \pi^2} \frac{2 \sqrt{2}}{F}  \epsilon_{\mu \nu \alpha \beta}
\end{split}
\\
\times \langle 2 \p_\mu \pi \p_\nu v_\alpha v_\beta - \p_\mu \pi v_\alpha  \p_\nu v_\beta \rangle
\end{multline*}
 We can check, that the longitudinal part is exactly the same as in  \cite{Vain}, if we, formally, expand it in $M^2/Q^2$:
\begin{multline*}
\langle A_{\parallel \mu}(-Q) V_{\nu}(Q) \tilde{V}_{\rho}(k) \rangle=\\
 =\langle T_A T_V T_{\tilde{V}} \rangle  \epsilon^{\parallel}_{\mu \nu \rho \sigma}(i{k}_{\sigma}) \frac{N_C}{2 \pi^2} \left[1 - \frac{M^2}{Q^2} \right]
\end{multline*}
($Q^2$ is Euclidean momentum)
\subsection{Relation with the Dirac operator spectrum}

Let us also briefly comment on the different calculation of the magnetic susceptibility
via the spectral density of the Dirac operator and introduce eigenfunctions of the
Dirac operator in the external gluon field A
\begin{equation*}
\hat{D}(A)u_{\lambda}(x)=\lambda u_{\lambda}(x)
\end{equation*}
Then the standard definition of the spectral density reads as
\begin{equation*}
\rho(\lambda)=\langle V^{-1}\sum_n \delta(\lambda -\lambda_n)\rangle_{A}
\end{equation*}
where $V$ is Euclidean volume and the averaging over the gluon ensemble
is assumed. The value of the spectral density at the origin is fixed
by the Casher-Banks relation \cite{casher} while the linear term
was determined comparing the different calculations of the correlator of the
scalar currents \cite{stern}. In the perturbation theory the spectral density
behaves as $\rho(\lambda)\propto \lambda^3$ that is starting from the third
order the universality is lost because of the mixing with the perturbative modes.
We would like to note that the magnetic susceptibility is sensitive to the last
"nonperturbative" quadratic $\lambda^2$ term in the spectral density.

To explain this point let us consider the " two-point loop diagram"
with tensor and vector vertexes in terms of the eigenfunctions
and eigenvalues of the Dirac operator.  The simple inspection shows
that the susceptibility is expressed in terms of  two different contributions.
The first "diagonal" contribution reads as
$$
m\int d\lambda  \frac{\rho(\lambda)}{(\lambda^2 +m^2)^2}
$$
while the second "nondiagonal" contributions involves the following integrals
$$
\int d^4 x \bar{u}_{\lambda}(x) x_{\nu} u_{\lambda'}(x)
$$
and double integrals over the eigenvalues $\int d\lambda \int d\lambda'$.
The "diagonal" contribution is IR divergent and this divergence is expected
to be  canceled by the "nondiagonal" terms amounting to a kind of sum rules.
On the other hand it is
clear that  quadratic term in the spectral density yields the finite
contribution. It is not clear if the "nondiagonal" terms yield the IR finite
contribution as well. This point does not allow us to write down
the coefficient in front of the $\lambda^2$ in the spectral density
immediately. One can not also exclude that more careful
treatment of the IR divergences  should involve the derivation a kind of
the effective action with the tensor insertion .
We hope to discuss these issues elsewhere.

\section {Conclusion}
In this paper we have derived the expression for the magnetic susceptibility
of the quark condensate in the holographic  QCD model. We have demonstrated
that this object captures  nontrivial anomalous properties of the dual model
encoded in the Chern-Simons term. It vanishes if the CS term is not taken into account.
The second important lesson concerns the validity of  Vainshtein's relation which
is not exact but is fulfilled with the high accuracy. 

The numerical value of the susceptibility do not coincide with recent estimations from the instanton liquid model \cite{dorokhov1,dorokhov2},sum rules fit \cite{Braun} and phenomenology of D-meson decays \cite{Rohr}. But it is calculated at significantly less energy scale: for our calculation $Q \ll 1150 \ \mathrm{Mev}$, while others are calculated at $Q \sim 1 \ \mathrm{Gev}$, so we do not find the contradiction. This also allows us to compare the result with Vainshtein, whose normalization point is about $0.5 \ \mathrm{Gev}$.

The only parameter the coefficient in Vainshtein's relation depends on is the IR cut-off scale however the dependence is very smooth.
It would be interesting to discuss the soft wall model and
derive the dependence of the magnetic susceptibility on  temperature
and  chemical potential.

\begin{acknowledgments}
We are grateful to I. Denisenko  and P. Kopnin for the useful discussions.
The work of A.G. was supported in part by grants,
INTAS-1000008-7865, PICS- 07-0292165 and of A.K.  by Russian President's Grant for Support of Scientific Schools NSh-3036.2008.2, by RFBR grant 09-02-00308 and Dynasty Foundation.
\end{acknowledgments}

\appendix*
\section{}
In Appendix, we state some results of \cite{Erlich,Krik} concerning the "hard wall" AdS/QCD model.
The 5D coupling constant $g_5$ is fixed by the 2-point function of vector currents in \cite{Erlich}
\be
\label{g_5}
\frac{g_5^2}{R^2} = \frac{12 \pi^2}{N_c}
\ee
The position of the IR boundary $z_m$ is related to the $\rho$-meson mass \cite{Erlich}.
\be
\label{z_m}
z_m = \frac{1}{323 Mev}
\ee
The parameter $\sigma$ is coupled with the value of quark condensate (we take the value $\langle \bar{q}q \rangle = (230 \mathrm{Mev})^3$) and equals \cite{Krik}:
\be
\label{sigma}
\sigma = \frac{N_f \langle \bar{q}q \rangle}{3 R^3 \Lambda^2} = (460 Mev)^3
\ee
We shall also fix the constant $\Lambda$  correcting calculation made in \cite{Krik}.
First,  compute  the leading order solutions to the equation of motion
for the pseudoscalar fields.
These are solutions to the equations of motion (\ref{9}), (\ref{9.5}) with fixed boundary value of $\phi$ at $z=\epsilon$. Differentiating (\ref{9.5}) and substituting $\partial_z \phi$ from (\ref{9}) we get :
\begin{multline*}
\partial_{z}^{2} \frac{v^2}{z^3} \partial_{z} \pi - \left(\partial_{z} \frac{v^2}{z^3}\right) \frac{z^3}{v^2} \partial_{z}\frac{v^2}{z^3} \partial_{z} \pi  \\
 -Q^2 \frac{v^2}{z^3} \partial_{z} \pi - \frac{g_{5}^{2} R^2 \Lambda^2 v^4}{z^5} \partial_{z} \pi = 0.
\end{multline*}
We need to solve it near the boundary, so  substitute   asymptotic value $v(z)=mz|_{z\rightarrow 0}$ and denoting $x=Qz$ it takes the form:
$$
\partial_{x}^{2} \frac{1}{x} \partial_{x} \pi + \frac{1}{x} \partial_{x} \frac{1}{x} \partial_{x} \pi - \frac{1}{x} \partial_{x} \pi - \frac{g_{5}^{2} R^2 \Lambda^2 m^2}{Q^2} \frac{1}{x} \partial_{x} \pi = 0.
$$
At large $Q^2$  we  neglect the last term and obtain the modified Bessel equation with $\lambda=0$ . Hence the solution for $\pi(z)$ reads as :
$$
\pi(z) = A' Q z I_{1}(Qz) + B' Q z K_{1} (Qz) - C'.
$$
and using (\ref{9})  we immediately obtain the solution for $\phi$:
$$
\phi(z) = - \frac{g_{5}^{2} R^2 \Lambda^2 m^2}{Q^2} Qz [A' I_{1}(Qz) + B' K_{1} (Qz)]+C'.
$$
The boundary condition on $\phi$ at $z=\epsilon$ fixes the constant $B'$:
$$
\phi(0,q) = \phi_{0} (q) = - \frac{g_{5}^{2} R^2 \Lambda^2 m^2}{Q^2}  B' + B'
$$
$$
B'=\frac{1}{1 - g_{5}^{2} R^2 \Lambda^2 \frac{m^2}{Q^2}} \phi_{0} (q)
$$
therefore we finally get :
\begin{align*}
\phi(z)|_{z=\epsilon}&=\phi_{0} (q) \\
\left. \frac{\partial_{z} \phi(z)}{z} \right|_{z=\epsilon}&= - \frac{g_{5}^{2} R^2 \Lambda^2 m^2}{Q^2} B' \phi_{0} (q) \frac{Q^2}{2} ln(Q^2 \epsilon^2) \notag\\
\pi(z)|_{z=\epsilon}&= 0
\end{align*}

We can compute the 2-point function of pseudoscalar currents, using the relation:
$$
\partial_{\mu}(\bar{q}\gamma_5 \gamma_{\mu} q ) = 2m_{q}(\bar{q}\gamma_5 q ).
$$
which yields us  the source for pseudoscalar current
$$
2m_{q} \phi \leftrightarrow (\bar{q}\gamma_5 q ).
$$

In order to obtain the 2-point function, we  vary the action twice with respect to $2m_{q} \phi(0)$ and find:

\begin{multline*}
\delta S_{\pi} =   \int d^{4} x \ \ \frac{R}{g_{5}^{2}} \left[ \delta \partial_{\mu} \phi \frac{\partial_z  \partial_{\mu} \phi }{z} \right]_{z=\epsilon}  - \Lambda^2 R^3 \left[ \delta \pi \frac{v^2}{z^3} \partial_z \pi \right]_{\epsilon} \\
= \int\limits_{(x,q_1,q_2)} e^{ \imath (q_{1}+q_{2}) x}
 \bigg( \frac{R Q^2}{g_{5}^{2}} \left[ \delta \phi(q_1,z) \frac{\partial_z  \phi(q_2,z) }{z} \right]_{z=\epsilon}  \\
  -\Lambda^2 R^3 m^2 \left[ \delta \pi(q_1,z) \frac{\partial_z \pi(q_2,z)}{z} \right]_{z=\epsilon} \bigg),
\end{multline*}

hence, the pseudoscalar correlator is:

\begin{multline*}
\langle J_{\pi}^{a} (q),J_{\pi}(q)^{b} \rangle =\\
=2 \delta^{ab} \frac{1}{4 m^2}\frac{R Q^2}{g_{5}^{2}} \left[ - \frac{g_{5}^{2} R^2 \Lambda^2 m^2}{Q^2} B' \frac{Q^2}{2} ln(Q^2 \epsilon^2)\right] \rightarrow \\
\xrightarrow[m=0]{} \frac{\delta^{ab}}{4 }{R^3 \Lambda^2} Q^2 ln(Q^2 \epsilon^2)
\end{multline*}

Comparing with the QCD value \cite{sumrule}:
\begin{align*}
\langle J_{\pi}^{a} (q),J_{\pi}(q)^{b} \rangle = \delta^{ab} \frac{N_c}{16 \pi^2} Q^2 ln(Q^2 \epsilon^2)
\end{align*}
we find
$$
\Lambda^2 R^3 = \frac{N_c}{4 \pi^2} = \frac{R}{3 g_5^2}
$$
\be
\label{k^2}
k^2 = R^2 \Lambda^2 g_5^2 = 3
\ee


\begin{thebibliography}{99}

\bibitem{Erlich} J.~Erlich, E.~Katz, D.~T.~Son and M.~A.~Stephanov,
  ``QCD and a Holographic Model of Hadrons,''
  Phys.\ Rev.\ Lett.\  {\bf 95}, 261602 (2005)
  [arXiv:hep-ph/0501128].

\bibitem{soft}
  A.~Karch, E.~Katz, D.~T.~Son and M.~A.~Stephanov,
  ``Linear Confinement and AdS/QCD,''
  Phys.\ Rev.\  D {\bf 74}, 015005 (2006)
  [arXiv:hep-ph/0602229].

\bibitem{Sugimoto}  T.~Sakai and S.~Sugimoto,
  ``Low energy hadron physics in holographic QCD,''
  Prog.\ Theor.\ Phys.\  {\bf 113}, 843 (2005)
  [arXiv:hep-th/0412141].



\bibitem{is}
 B.~L.~Ioffe and A.~V.~Smilga,
  ``Nucleon Magnetic Moments And Magnetic Properties Of Vacuum In QCD,''
  Nucl.\ Phys.\  B {\bf 232}, 109 (1984).

\bibitem{bal}
  I.~I.~Balitsky and A.~V.~Yung,
  ``Proton And Neutron Magnetic Moments From QCD Sum Rules,''
  Phys.\ Lett.\  B {\bf 129} (1983) 328.\\
  I.~I.~Balitsky, A.~V.~Kolesnichenko and A.~V.~Yung,
  ``On Vector Dominance In Sum Rules For Electromagnetic Hadron
  Characteristics. (In Russian),''
  Sov.\ J.\ Nucl.\ Phys.\  {\bf 41} (1985) 178
  [Yad.\ Fiz.\  {\bf 41} (1985) 282].
\bibitem{bel}
  V.~M.~Belyaev and Y.~I.~Kogan,
  ``Calculation Of Quark Condensate Magnetic Susceptibility By QCD Sum Rule
  Method,''
  Yad.\ Fiz.\  {\bf 40}, 1035 (1984).




\bibitem{Vain}
 A.~Vainshtein,
  ``Perturbative and nonperturbative renormalization of anomalous quark
  triangles,''
  Phys.\ Lett.\  B {\bf 569}, 187 (2003)
  [arXiv:hep-ph/0212231].

\bibitem{Pomarol} L.~Da Rold and A.~Pomarol,
  ``Chiral symmetry breaking from five dimensional spaces,''
  Nucl.\ Phys.\  B {\bf 721}, 79 (2005)
  [arXiv:hep-ph/0501218].


\bibitem{Krik}  A.~Krikun,
  ``On two-point correlation functions in AdS/QCD,''
  Phys.\ Rev.\  D {\bf 77}, 126014 (2008)
  [arXiv:0801.4215 [hep-th]].

\bibitem{Rad1}   H.~R.~Grigoryan and A.~V.~Radyushkin,
  ``Structure of Vector Mesons in Holographic Model with Linear Confinement,''
  Phys.\ Rev.\  D {\bf 76}, 095007 (2007)
  [arXiv:0706.1543 [hep-ph]].

\bibitem{Rad2}   H.~R.~Grigoryan and A.~V.~Radyushkin,
  ``Anomalous Form Factor of the Neutral Pion in Extended AdS/QCD Model with
  Chern-Simons Term,''
  Phys.\ Rev.\  D {\bf 77}, 115024 (2008)
  [arXiv:0803.1143 [hep-ph]].

\bibitem{sumrule} M.~A.~Shifman, A.~I.~Vainshtein and V.~I.~Zakharov,
  ``QCD And Resonance Physics. Sum Rules,''
  Nucl.\ Phys.\  B {\bf 147}, 385 (1979).\\
 V.~A.~Novikov, M.~A.~Shifman, A.~I.~Vainshtein and V.~I.~Zakharov,
  ``Are All Hadrons Alike?,''
  Nucl.\ Phys.\  B {\bf 191}, 301 (1981).

\bibitem{ChPT}
   J.~Gasser and H.~Leutwyler,
  ``Chiral Perturbation Theory To One Loop,''
  Annals Phys.\  {\bf 158}, 142 (1984).



\bibitem{casher}
  T.~Banks and A.~Casher,
  ``Chiral Symmetry Breaking In Confining Theories,''
  Nucl.\ Phys.\  B {\bf 169}, 103 (1980).

\bibitem{stern}
  A.~V.~Smilga and J.~Stern,
  ``On The Spectral Density Of Euclidean Dirac Operator In QCD,''
  Phys.\ Lett.\  B {\bf 318}, 531 (1993).

\bibitem{dorokhov1}
  A.~E.~Dorokhov,
  ``V A V-tilde correlator within the instanton vacuum model,''
  Eur.\ Phys.\ J.\  C {\bf 42} (2005) 309
  [arXiv:hep-ph/0505007].


\bibitem{dorokhov2}
    A.~E.~Dorokhov,
 ``Singlet V A V-tilde correlator within the instanton vacuum model,''
 JETP Lett.\  {\bf 82} (2005) 1
 [Pisma Zh.\ Eksp.\ Teor.\ Fiz.\  {\bf 82} (2005) 3]
 [arXiv:hep-ph/0505196].

\bibitem{Braun}
    P.~Ball, V.~M.~Braun and N.~Kivel,
 ``Photon distribution amplitudes in QCD,''
 Nucl.\ Phys.\  B {\bf 649} (2003) 263
 [arXiv:hep-ph/0207307].

\bibitem{Rohr}
 J.~Rohrwild,
 ``Determination of the magnetic susceptibility of the quark condensate using
 radiative heavy meson decays,''
 JHEP {\bf 0709} (2007) 073
 [arXiv:0708.1405 [hep-ph]].

\bibitem{Westenberger}
 J.~Erlich, C.~Westenberger,``Tests of Universality in AdS/QCD'' [arXiv:0812.5105 [hep-ph]].

\bibitem{Polchinski}
 J.~Polchinski, M.~J.~Strassler,
``Hard scattering and gauge / string duality'',
 Phys.Rev.Lett.88:031601,2002
[arXiv: hep-th/0109174]

\bibitem{Boschi1}
 H.~Boschi-Filho, N.~R.~F.~Braga,
 ``QCD / string holographic mapping and glueball mass spectrum'',
 Eur.Phys.J.C32:529-533,2004
 [arXiv: hep-th/0209080]

\bibitem{Boschi2}
  H.~Boschi-Filho, N.~R.~F.~Braga,
 ``Gauge / string duality and scalar glueball mass ratios'', JHEP 0305:009,2003
 [arXiv: hep-th/0212207]

\end{thebibliography}
\end{document}